\documentclass[11pt, pdftex,preprint]{elsarticle}
\usepackage[usenames,dvipsnames,svgnames,table]{xcolor}
\usepackage[colorlinks]{hyperref}
\AtBeginDocument{\hypersetup{linkcolor=blue, citecolor=OrangeRed}}
\usepackage{bm, bbm, bbding}
\usepackage{seqsplit}
\usepackage{amssymb,amsmath,amsfonts,mathrsfs,graphicx,caption,kantlipsum,units,yfonts}
\usepackage{txfonts,dsfont}
\usepackage[toc,page,title,titletoc,header]{appendix}
\usepackage{lineno,setspace, framed, hyperref}
\usepackage[top=2.6cm, bottom=3cm, left=3cm, right=3cm]{geometry}
\usepackage[novbox]{pdfsync}

\usepackage[numbers]{natbib}

\makeatletter
\def\ps@pprintTitle{
  \let\@oddhead\@empty
  \let\@evenhead\@empty
  \def\@oddfoot{\reset@font\hfil\thepage\hfil}
  \let\@evenfoot\@oddfoot}
\makeatother

\journal{}
\begin{document}
\begin{frontmatter}

\title{Local energy-momentum conservation in scalar-tensor-like gravity with generic curvature invariants}

\author{David Wenjie Tian\fnref{myfootnote}}
\address{Faculty of Science,  Memorial University, St. John's, Newfoundland, A1C 5S7, Canada}
\fntext[myfootnote]{Email address: wtian@mun.ca}

\begin{abstract}

For a large class of scalar-tensor-like gravity whose action contains nonminimal couplings between a scalar field $\phi(x^\alpha)$ and generic curvature invariants $\left\{\mathcal{R}\right\}$ beyond the Ricci scalar $R=R^\alpha_{\;\;\alpha}$, we prove the covariant invariance of its field equation and confirm/prove the local energy-momentum conservation. These $\phi(x^\alpha)-\mathcal{R}$ coupling terms break the symmetry of diffeomorphism invariance under a particle transformation, which implies
that the solutions to the field equation should satisfy the consistency condition $\mathcal{R}\equiv 0$ when  $\phi(x^\alpha)$ is nondynamical and massless. Following this fact and based on the accelerated expansion of the observable Universe, we propose a primary test to check the viability of the modified gravity to be an effective dark energy, and a simplest example passing the test is the ``Weyl/conformal dark energy''.\\


\end{abstract}

\end{frontmatter}

\section{Introduction}\label{Sec Introduction}

An important problem in relativistic theories of gravity is the divergence-freeness of the field equation and the covariant conservation of the energy-momentum tensor. In general relativity (GR), Einstein's equation $G_{\mu\nu}\equiv R_{\mu\nu}-\frac{1}{2}Rg_{\mu\nu}=8\pi G T_{\mu\nu}^{(m)}$ has a vanishing covariant divergence due to the contracted Bianchi identities $\nabla^\mu G_{\mu\nu}\equiv 0$, which guarantees the local energy-momentum conservation $\nabla^\mu T_{\mu\nu}^{(m)}=0$ or $\partial^{\,\mu}(\!\sqrt{-g}\,T_{\mu\nu}^{(m)})=0$ for the matter tensor $T_{\mu\nu}^{(m)}$. In modified gravities beyond GR and its Hilbert-Einstein action, the conservation problem becomes more complicated and has attracted a lot of interest.

In Ref.\cite{EMC GBI Palatini f(R)}, the generalized Bianchi identities were derived for the Palatini formulation of the nonlinear $f(R)$ gravity, and its local energy-momentum conservation was further confirmed in
Ref.\cite{EMC Palatini L(R)} by the equivalence between Palatini $f(R)$ and the $\omega=-3/2$ Brans-Dicke gravity. Ref.\cite{EMC Palatini f(R phi)} investigated a mixture of $f(R)$ and the generalized Brans-Dicke gravity, and proved the covariant conservation from both the metric and the Palatini variational approaches. For Einstein-Cartan gravity which allows for spacetime torsion, both the energy-momentum and the angular momentum conservation were studied in Ref.\cite{EMC Einstein Cartan} by decomposing the Bianchi identities in Riemann-Cartan spacetimes. In Refs.\cite{EMC Palatini f(R phi), Nonminimal Coupling I, Nonminimal Coupling II, AA Tian Paper}, the nontrivial divergences $\nabla^\mu T_{\mu\nu}^{(m)}$ were analyzed for the situations where the matter Lagrangian density is multiplied by different types of curvature invariants in the action. Also, interestingly in Ref.\cite{EMC nonconservation}, the possible consequences after dropping the energy-momentum conservation in GR, such as the modified evolution equation for the Hubble parameter, were investigated.

Besides the covariant invariance $\nabla^\mu T_{\mu\nu}^{(m)}=0$ for the matter tensor $T_{\mu\nu}^{(m)}$ that has been standardly defined in GR and modified gravities (cf. Eq.(\ref{continuity conservation I}) below), the conservation problem has also been studied for more fundamental definitions of energy-momentum tensors from a wider perspective, i.e. from a first-principle approach making use of Noether's theorem and the classical field theory. For example,  the Noether-induced canonical energy-momentum conservation for the translational invariance of the Lagrangian was studied in Ref.\cite{General conservation I} for general spacetimes with torsion and nonmetricity. The conservation equations and the Noether currents for the Poincar\'e-transformation  invariance were studied in Ref.\cite{General conservation II} for the 3+1 and 2+1 dimensional Einstein gravity and the 1+1 dimensional string-inspired gravity. Also, Refs.\cite{EMC metric-torsion I} and \cite{EMC metric-torsion II} extensively discussed the diffeomorphically invariant metric-torsion gravity whose action contains first- and second-order derives of the torsion tensor, and derived the full set of Klein-Noether differential identities and various types of conserved currents.

In this paper, our interest is the covariant invariance of such modified gravities whose actions involve nonminimal couplings between arbitrary curvature invariants $\left\{\mathcal{R}\right\}$ and a background scalar field $\phi(x^\alpha)$. For example, $\phi(x^\alpha)$ is coupled to the Ricci scalar $R=R^\alpha_{\;\;\alpha}$ in Brans-Dicke and scalar-tensor gravity in the Jordan frame \cite{Brans Dicke}, to the Chern-Pontryagin topological density in the Chern-Simons modification of GR \cite{Chern-Simons 1}, and to the Gauss-Bonnet invariant $\mathcal{G} = R^2-4R_{\alpha\beta}R^{\alpha\beta}+R_{\alpha\mu\beta\nu}R^{\alpha\mu\beta\nu}$ in the Gauss-Bonnet effective dark energy \cite{GaussBonnet dark energy}. In theory, one could consider the nonminimal coupling of $\phi(x^\alpha)$ to an arbitrarily complicated curvature invariant beyond the Ricci scalar. In such situations, however, the covariant invariance of the field equation has not been well understood, so we aim to carefully look into this problem by this work. Note that it might sound more complete to analyze the global conservation $\nabla^\mu (T_{\mu\nu}^{(m)}+t_{\mu\nu})=0$ or $\partial^{\,\mu}[\!\!\sqrt{-g}(T_{\mu\nu}^{(m)}+t_{\mu\nu})]=0$, where $t_{\mu\nu}$ refers to the energy-momentum pseudotensor for the gravitational field, but to make this paper more clear and readable, we choose to concentrate on the local conservation $\nabla^\mu T_{\mu\nu}^{(m)}=0$, while the incorporation of $t_{\mu\nu}$ will be discussed separately.

This paper is organized as follows. In Sec.~\ref{Sec General formulation}, we introduce the generic class of modified gravity with the nonminimal $\phi(x^\alpha)-$couplings to arbitrary Riemannian invariants $\left\{\mathcal{R}\right\}$, calculate the divergence for different parts of the total action, prove the covariant invariance of the field equation, and confirm the local energy-momentum conservation. Section~\ref{Sec Nondynamical scalar field} investigates the reduced situations that the scalar field is nondynamical and massless, and derives the consistency constraint  $\mathcal{R}\equiv 0$ which suppresses the breakdown of diffeomorphism invariance. Finally, applications of the theories in Secs.~\ref{Sec General formulation} and ~\ref{Sec Nondynamical scalar field} are considered in Sec.~\ref{Sec Applications}. Throughout this paper,  we adopt the geometric conventions $\Gamma^\alpha_{\beta\gamma}=\Gamma^\alpha_{\,\;\;\beta\gamma}$, $R^{\alpha}_{\;\;\beta\gamma\delta}=
\partial_\gamma \Gamma^\alpha_{\delta\beta}-\partial_\delta \Gamma^\alpha_{\gamma\beta}\cdots$ and 
$R_{\mu\nu}=R^\alpha_{\;\;\mu\alpha\nu}$ with the metric signature $(-,+++)$.

\section{General theory}\label{Sec General formulation}

\subsection{Scalar-tensor-like gravity}\label{Sec Scalar-tensor-like gravity}

Consider a theory of modified gravity or effective dark energy given by the following action,
\begin{eqnarray}\label{Generic action}
\mathcal{S}=\int d^4x \sqrt{-g}\,\left(\mathscr{L}_{\text{HE}}+\mathscr{L}_{\text G}
+\mathscr{L}_{\text{NC}}+\mathscr{L}_{\phi} \right)+\mathcal{S}_m\,,
\end{eqnarray}
where $\mathscr{L}_{\text{HE}}$ refers to the customary Hilbert-Einstein Lagrangian density as in GR,
\begin{eqnarray}
\mathscr{L}_{\text{HE}}= R\,,
\end{eqnarray}
while $\mathscr{L}_{\text G}$ denotes the extended dependence on generic curvature invariants $\mathcal{R}$,
\begin{eqnarray}
\mathscr{L}_{\text G}=  f\left(R,\cdots, \mathcal{R}  \right)\,.
\end{eqnarray}
Here $\mathcal{R} =\mathcal{R} \,\big(g_{\alpha\beta}\,,R_{\mu\alpha\nu\beta}\,,
\nabla_\gamma R_{\mu\alpha\nu\beta}\,,\ldots\big)$
is an arbitrary invariant function of the metric as well as the Riemann tensor and its derivatives up to any order. For example, $\mathcal{R}$ can come from the fourteen\footnote{When one combines the spacetime geometry with matter fields in the framework of GR, the amount of independent algebraic invariants will be extended to sixteen in the presence of electromagnetic or perfect-fluid fields \cite{Riemann invariants II}.} algebraically independent real invariants of the Riemann tensor \cite{Riemann invariants} and their combinations, say  $R_{\alpha\beta}R^{\alpha\beta}+R_{\alpha\mu\beta\nu}R^{\alpha\mu\beta\nu}
+R_{\alpha\mu\beta\nu}R^{\alpha\beta}R^{\mu\nu}$,  which will yield fourth--order field equations; or differential Riemannian invariants that will lead to sixth-- or even higher--order field equations, like $R\nabla^\alpha\nabla_\alpha R+R^\alpha_{\;\;\mu}R^\beta_{\;\;\nu}\nabla_\alpha\nabla_\beta R^{\mu\nu}$.

In the total Lagrangian density, $\mathscr{L}_{\text{NC}}$ represents the nonminimal coupling effects,
\begin{eqnarray}
\mathscr{L}_{\text{NC}}= h(\phi)\cdot \widehat{f}\left(R,\cdots, \mathcal{R}  \right) \,,
\end{eqnarray}
where $h(\phi)$ is an arbitrary function of the scalar field $\phi=\phi(x^\alpha)$, and $ \widehat{f}\left(R,\cdots, \mathcal{R}  \right)$ has generic dependence on curvature invariants, with the dots ``$\cdots$'' in $\widehat{f}\left(R,\cdots, \mathcal{R}  \right)$ and the $ f\left(R,\cdots, \mathcal{R}  \right)$ above denoting different choices of $\mathcal{R}$. Moreover, the kinetics of  $\phi(x^\alpha)$ is governed by
\begin{eqnarray}\label{Lagrangian phi}
\mathscr{L}_{\phi} = -\lambda(\phi)\cdot\nabla_\alpha\phi\nabla^\alpha\phi
-V(\phi)\,.
\end{eqnarray}
In the  $(-,+++)$ system of conventions, $\phi(x^\alpha)$ is canonical if  $\lambda(\phi)>0$, noncanonical if $\lambda(\phi)<0$, and nondynamical if $\lambda(\phi)=0$.

Finally, as usual, the matter action $\mathcal{S}_m$ in Eq.(\ref{Generic action}) is given by the matter Lagrangian density via
\begin{eqnarray}\label{Sm}
\mathcal{S}_m=16\pi G \int d^4x \sqrt{-g}\,\mathscr{L}_m\left(g_{\mu\nu}, \psi_m, \partial_\mu\psi_m\right)\,,
\end{eqnarray}
where the variable $\psi_m$ collectively describes the matter fields, and $\psi_m$ is minimally coupled to the metric tensor $g_{\mu\nu}$. Unlike the usual dependence on $\partial_\mu\psi_m$ in its standard form, $\mathscr{L}_m=\mathscr{L}_m\left(g_{\mu\nu}, \psi_m, \partial_\mu\psi_m\right)$ does not contain derivatives of the metric tensor -- such as Christoffel symbols or curvature invariants, in light of the minimal gravity-matter coupling and Einstein's equivalence principle; physically, this means $\mathscr{L}_m$ reduces to the matter Lagrangian density for the flat spacetime in a freely falling local reference frame (i.e. a locally geodesic coordinate system).

To sum up, we are considering the modifications of GR  into the total Lagrangian density $\mathscr{L}=R+f\left(R,\cdots, \mathcal{R}  \right)+h(\phi)\cdot \widehat{f}\left(R,\cdots, \mathcal{R}  \right) -\lambda(\phi)\cdot\nabla_\alpha\phi\nabla^\alpha\phi  -V(\phi)+16\pi G \mathscr{L}_m$, which has been rescaled so that the numerical coefficient $16\pi G$ is associated to $\mathscr{L}_m$.
It can be regarded as a mixture of the nonlinear higher-order gravity  $\mathscr{L} =R+f\left(R,\cdots, \mathcal{R}  \right)+16\pi G \mathscr{L}_m$ in the metric formulation for the curvature invariants, and the generalized scalar-tensor gravity  $\mathscr{L} =h(\phi)\cdot \widehat{f}\left(R,\cdots, \mathcal{R}  \right) -\lambda(\phi)\cdot\nabla_\alpha\phi\nabla^\alpha\phi  -V(\phi)+16\pi G \mathscr{L}_m$ in the Jordan frame for the scalar field. Hereafter we will mainly work with actions,  and for simplicity we will sometimes adopt the total Lagrangian density in place of the corresponding action in full integral form.

\subsection{Divergence-freeness of gravitational field equation}\label{Sec Divergence of field equation}
\vspace{3mm}

\subsubsection{Pure curvature parts}
\vspace{3mm}

For the Hilbert-Einstein part of the total action, i.e.  $\mathcal{S}_{\text{HE}}= \int d^4x \sqrt{-g}\, \mathscr{L}_{\text{HE}}$, its variation with respect to the inverse metric yields the well-known result $\delta\mathcal{S}_{\text{HE}}\cong \int d^4x \sqrt{-g}\, G_{\mu\nu}\delta g^{\mu\nu}$. By the symbol $\cong$ we mean the equality after neglecting all total derivatives in the integrand or equivalently boundary terms of the action when integrating by parts, and the Einstein tensor $G_{\mu\nu}=R_{\mu\nu}-\frac{1}{2}Rg_{\mu\nu}$ respects the twice-contracted Bianchi identity $\nabla^\mu G_{\mu\nu}=0$.

For the gravitational action $\mathcal{S}_{\text G}= \int d^4x \sqrt{-g}\, \mathscr{L}_{\text G}$ for the extended dependence on generic Riemannian invariants, formally we write down the variation as $\delta\mathcal{S}_{\text G} \cong  \int d^4x \sqrt{-g}\, H_{\mu\nu}^{\text{(G)}}\delta g^{\mu\nu}$,  where $H_{\mu\nu}^{\text{(G)}}$ resembles and generalizes the Einstein tensor by
\begin{equation}
H_{\mu\nu}^{\text{(G)}}\,\cong\,\frac{1}{\sqrt{-g}} \,\frac{\delta\, \left[\!\!\sqrt{-g}\,f(R,\cdots,\mathcal{R})\right]}{\delta g^{\mu\nu}}\,.
\end{equation}
Due to the coordinate invariance of $\mathcal{S}_{\text G}$,  $H_{\mu\nu}^{\text{(G)}}$ satisfies the generalized contracted Bianchi identities \cite{Eddington Generalized Bianchi and Conservation, BDfR equivalence Generalized Bianchi}
\begin{equation}\label{Generalized contracted Bianchi identities}
\nabla^\mu \,\left( \frac{1}{\sqrt{-g}} \,\frac{\delta\,
\left[\!\!\sqrt{-g}\,f(R,\cdots,\mathcal{R})\right]}{\delta g^{\mu\nu}}\right)=\,0\,,
\end{equation}
or just $\nabla^\mu H_{\mu\nu}^{\text{(G)}}=0$ by the definition of  $H_{\mu\nu}^{\text{(G)}}$. Similar to  the relation  $G_{\mu\nu}=R_{\mu\nu}-\frac{1}{2}Rg_{\mu\nu}$,   one can further expand $H_{\mu\nu}^{\text{(G)}}$ to rewrite Eq.(\ref{Generalized contracted Bianchi identities})  into
\begin{equation}
\nabla^\mu \,\left( f_R R_{\mu\nu}+
\sum f_{\mathcal{R} }  \mathcal{R}_{\mu\nu} -\frac{1}{2}f(R,\cdots,\mathcal{R}) \,g_{\mu\nu}\right)=0\,,
\end{equation}
where $f_R \coloneqq \partial f(R ,\cdots,\mathcal{R})/\partial R$, $f_{\mathcal{R} } \coloneqq \partial  f(R,\cdots,\mathcal{R})/\partial \mathcal{R} $, and $\mathcal{R}_{\mu\nu}\cong
\left(f_{\mathcal{R} } \delta \mathcal{R}  \right)/\delta g^{\mu\nu}$ -- note that in the calculation of  $\mathcal{R}_{\mu\nu}$, $f_{\mathcal{R} }$ will serve as a nontrivial coefficient if
$f_{\mathcal{R} }\neq \text{constant}$ and should be absorbed into the variation $ \delta \mathcal{R}$ when integrated by parts.

\subsubsection{Nonminimal $\phi(x^\alpha)$-curvature coupling part}
\vspace{3mm}

For the componential action $\mathcal{S}_{\text{NC}}= \int d^4x \sqrt{-g}\, \mathscr{L}_{\text{NC}}$ for the nonminimal coupling effect, formally we have  the variation $\delta\mathcal{S}_{\text{NC}} \cong  \int d^4x \sqrt{-g}\,  H_{\mu\nu}^{\text{(NC)}}\delta g^{\mu\nu}$, where
\begin{equation}
H_{\mu\nu}^{\text{(NC)}}\,\cong\,\frac{1}{\sqrt{-g}}\frac{\delta\left[\!\!\sqrt{-g}\,\phi\,\widehat{f}\left(R,\cdots, \mathcal{R}  \right)\right]}{\delta g^{\mu\nu}}\,.
\end{equation}
Unlike $H_{\mu\nu}^{\text{(G)}}$ with Eq.(\ref{Generalized contracted Bianchi identities}), $H_{\mu\nu}^{\text{(NC)}}$ does not respect some straightforward generalized Bianchi identities; this is because $\mathcal{S}_{\text{NC}}$ involves the coupling with the background scalar field $\phi(x^\alpha)$ and is no longer purely tensorial gravity. Thus, we will analyze the divergence of $H_{\mu\nu}^{\text{(NC)}}$ by the diffeomorphism of $\mathcal{S}_{\text{NC}}$.

Consider an arbitrary infinitesimal coordinate transformation $x^\mu\mapsto x^\mu+\delta x^\mu$, where $\delta x^\mu= k^\mu$ is an infinitesimal vector field that vanishes on the boundary, $\left.k^\mu=0\,\right|_{\partial\Omega}$, so that the spacetime manifold is mapped onto itself. $\mathcal{S}_{\text{NC}}$ responds to this transformation by
\begin{eqnarray}
\delta \mathcal{S}_{\text{NC}}=\!\!&-&\!\!\int d^4x \,h(\phi)\cdot\partial_\mu \left[ k^\mu\sqrt{-g} \,\widehat{f}\left(R,\cdots, \mathcal{R}  \right) \right]\label{conservation constraint step 5}\\
\cong\!\!&&\!\!\int d^4x\sqrt{-g} \,\widehat{f}\left(R,\cdots, \mathcal{R}  \right) \cdot \left(h_\phi\partial_\mu\phi \right) k^\mu, \label{conservation constraint step 6}
\end{eqnarray}
where $h_\phi\coloneqq dh(\phi)/d\phi$. For Eq.(\ref{conservation constraint step 5}), one should note that $\phi(x^\alpha)$ acts as a fixed background, as it only relies on the coordinates (i.e. spatial location and time) and is independent of the spacetime metric; moreover, the coordinate shift $x^\mu\mapsto x^\mu+k^\mu$ is a  particle/active transformation,
under which the dynamical tensor field $g_{\mu\nu}$ and thus $\sqrt{-g} \,\widehat{f}\left(R,\cdots, \mathcal{R}  \right)$ transform, while the background field $\phi(x^\alpha)$ and the coordinate system parameterizing the spacetime remain unaffected \cite{Diffeomorphism Explicit Spontaneous}.

Under the particle transformation $x^\mu\mapsto x^\mu+k^\mu$, the metric tensor varies by $ g_{\mu\nu}\mapsto  g_{\mu\nu} +\delta g_{\mu\nu} $ with $\delta g_{\mu\nu}=$ $\pounds_{\vec k} g_{\mu\nu}  =-\nabla_\mu k_\nu-\nabla_\nu k_\mu$, and therefore $ g^{\mu\nu}\mapsto  g^{\mu\nu} +\delta g^{\mu\nu} $ with $\delta g^{\mu\nu}=-\pounds_{\vec k} g^{\mu\nu}=\nabla^\mu k^\nu+\nabla^\nu k^\mu$. Recalling the definition of  $H_{\mu\nu}^{\text{(NC)}}$ with $H_{\mu\nu}^{\text{(NC)}}$ being symmetric for the index switch $\mu\leftrightarrow\nu$, one has
\begin{equation}
\begin{split}
\delta \mathcal{S}_{\text{NC}}
=\, 2\int d^4x \sqrt{-g}\,  H_{\mu\nu}^{\text{(NC)}} \nabla^\mu k^\nu \,
\cong  -2\int d^4x\sqrt{-g}\, \left(\nabla^\mu
H_{\mu\nu}^{\text{(NC)}}\right) k^\nu,\label{conservation constraint step 7}
\end{split}
\end{equation}
Comparing Eq.(\ref{conservation constraint step 6}) with Eq.(\ref{conservation constraint step 7}), we conclude that $H_{\mu\nu}^{\text{(NC)}}$ has a nontrivial divergence for $\phi(x^\alpha)\neq\text{constant}$, and
\begin{equation}\label{Divergence NC}
\nabla^\mu  H_{\mu\nu}^{\text{(NC)}}
=-\frac{1}{2}\,\widehat{f}\left(R,\cdots, \mathcal{R}  \right) \cdot h_\phi\partial_\nu \phi \,.
\end{equation}
In fact, Eq.(\ref{Divergence NC}) reflects the breakdown of diffeomorphism invariance in the presence of a fixed background scalar field.
As a comparison, it is worthwhile to mention that under an observer/passive transformation where the observer or equivalently the coordinate system transforms, both the tensor fields and the background scalar field will be left unchanged, so the symmetry of observer-transformation invariance continues to hold \cite{Diffeomorphism Explicit Spontaneous}.


\subsubsection{Purely scalar-field part}
\vspace{3mm}

Next, for the purely scalar-field part $\mathcal{S}_{\phi}= \int d^4x \sqrt{-g}\, \mathscr{L}_{\phi}$ with the variation $\delta\mathcal{S}_{\phi} \cong\int d^4x \sqrt{-g}\, H_{\mu\nu}^{(\phi)}\delta g^{\mu\nu}$, explicit calculations find
\begin{equation}\label{Hmunu phi}
\begin{split}
H_{\mu\nu}^{(\phi)}=-\lambda(\phi)\cdot\nabla_\mu\phi\nabla_\nu\phi
+\frac{1}{2}\bigg(\lambda(\phi)\cdot\nabla_\alpha\phi\nabla^\alpha\phi+V(\phi)\bigg)\,g_{\mu\nu}\,.
\end{split}
\end{equation}
Taking its contravariant derivative, we immediately obtain the nontrivial divergence
\begin{equation}\label{Divergence phi 1}
\begin{split}
\nabla^\mu H_{\mu\nu}^{(\phi)}=
-\frac{1}{2}\left(\lambda_\phi\cdot\nabla_\alpha\phi\nabla^\alpha\phi
+2\lambda(\phi)\cdot\Box\phi -V_\phi\right)\cdot\nabla_\nu \phi\,,
\end{split}
\end{equation}
where $\lambda_\phi\coloneqq d\lambda(\phi)/d\phi$, $V_\phi\coloneqq dV(\phi)/d\phi$, and $\Box$ denotes the covariant d'Alembertian with $\Box\phi=g^{\alpha\beta}\nabla_\alpha\nabla_\beta\phi
=\frac{1}{\sqrt{-g}}
\partial_\alpha (\!\sqrt{-g}\,g^{\alpha\beta}\partial_\beta \phi )$.
On the other hand, extremizing the entire action Eq.(\ref{Generic action}) with respect to the scalar field, i.e. $\delta S/\delta \phi=0$, one could obtain the kinematical wave equation
\begin{equation}\label{phi wave 1}
\begin{split}
2\lambda(\phi)\cdot\Box\phi= -\widehat{f}\left(R,\cdots, \mathcal{R}  \right) \cdot h_\phi- \lambda_\phi\cdot\nabla_\alpha\phi\nabla^\alpha\phi +V_\phi\,.
\end{split}
\end{equation}
We regard it as ``kinematical''
because it does not explicitly relate the propagation of $\phi(x^\alpha)$
to $T^{\text{(m)}}=g^{\mu\nu}T^{\text{(m)}}_{\mu\nu}$ for the matter distribution, while the ``dynamical'' wave equation can be obtained after combing Eq.(\ref{phi wave 1}) with the trace of the gravitational field equation.
Substitute Eq.(\ref{phi wave 1}) into the right hand side of Eq.(\ref{Divergence phi 1}), and it follows that
\begin{equation}\label{Divergence phi 2}
\begin{split}
\nabla^\mu H_{\mu\nu}^{(\phi)}= \frac{1}{2}\widehat{f}\left(R,\cdots, \mathcal{R}  \right) \cdot h_\phi\nabla_\nu\phi \,,
\end{split}
\end{equation}
which exactly cancels out the divergence of  $H_{\mu\nu}^{\text{(NC)}}$ in Eq.(\ref{Divergence NC}) for the nonminimal-coupling part $\mathcal{S}_{\text{NC}}$.

\subsubsection{Covariant invariance of field equation and local energy-momentum conservation}
\vspace{3mm}

To sum up, for the modified gravity or effective dark energy given by Eq.(\ref{Generic action}), its field equation reads
\begin{equation}\label{Field eqn}
\begin{split}
G_{\mu\nu}+H_{\mu\nu}^{\text{(G)}}+H_{\mu\nu}^{\text{(NC)}}+H_{\mu\nu}^{(\phi)}= 8\pi GT_{\mu\nu}^{(m)} \,,
\end{split}
\end{equation}
where, unlike $G_{\mu\nu}$ and $H_{\mu\nu}^{(\phi)}$, the exact forms of $\left\{H_{\mu\nu}^{\text{(G)}}\right.$,  $\left.H_{\mu\nu}^{\text{(NC)}}\right\}$ will not be determined until the concrete expressions of $\left\{\mathscr{L}_{\text G}\right.$, $\left.\mathscr{L}_{\text{NC}}\right\}$ or $\left\{f(R,\cdots,\mathcal{R})\,, \widehat{f}(R,\cdots,\mathcal{R})  \right\}$ are set up. In Eq.(\ref{Field eqn}), the energy-momentum tensor $T^{\text{(m)}}_{\mu\nu}$ is defined as in GR via \cite{Convention MTW Gravitation}
\begin{equation}\label{continuity conservation I}
\begin{split}
\delta \mathcal{S}_m   =-\frac{1}{2}\times 16\pi G \int  d^4x \sqrt{-g}\,T_{\mu\nu}^{\text{(m)}} \delta g^{\mu\nu}\quad\text{ with }\quad
T_{\mu\nu}^{\text{(m)}}\,\coloneqq\, \frac{-2}{\sqrt{-g}} \,\frac{\delta\, \Big(\!\!\sqrt{-g}\,\mathscr{L}_m \Big)}{\delta g^{\mu\nu}}\,,
\end{split}
\end{equation}
with $\mathcal{S}_m $ rescaled by $16\pi G$ in Eq.(\ref{Sm}). Instead of the variational definition Eq.(\ref{continuity conservation I}),  it had been suggested that $T^{\text{(m)}}_{\mu\nu}$ could be derived solely from the equations of motion $\displaystyle \frac{\partial \mathscr{L}_m}{\partial \psi_m}-\nabla_\mu\frac{\partial \mathscr{L}_m}{\partial (\partial_\mu\psi_m)}=0$ for the $\psi_m$ field in $\mathscr{L}_m\left(g_{\mu\nu}, \psi_m, \partial_\mu\psi_m\right)$ \cite{EMC tensor from matter function}; however, further analyses have shown that
this method does not hold a general validity, and Eq.(\ref{continuity conservation I}) remains as the most reliable approach to  $T^{\text{(m)}}_{\mu\nu}$ \cite{EMC from matter function}.

Adding up the (generalized) contracted Bianchi identities $\nabla^\mu G_{\mu\nu}=0$ and Eq.(\ref{Generalized contracted Bianchi identities}), and the nontrivial divergences Eqs.(\ref{Divergence NC}) and (\ref{Divergence phi 2}), eventually we conclude that the left hand side of the field equation (\ref{Field eqn}) is divergence free, the local energy-momentum conservation $\nabla^\mu T_{\mu\nu}^{(m)}=0$ holds, and the tensorial equations of motion for test particles remain the same as in GR.

In fact, the matter Lagrangian density $\mathscr{L}_m=\mathscr{L}_m\left(g_{\mu\nu}, \psi_m, \partial_\mu\psi_m\right)$ is a scalar invariant that respects the diffeomorphism invariance under the particle transformation $x^\mu\mapsto x^\mu+k^\mu$, and Noether's conservation law directly yields
\begin{equation}\label{continuity conservation II}
\nabla^\mu \,\left( \frac{1}{\sqrt{-g}} \,\frac{\delta\,
\Big(\!\!\sqrt{-g}\,\mathscr{L}_m \Big)}{\delta g^{\mu\nu}}\right)=\,0\,,
\end{equation}
which can be recast into $\displaystyle -\frac{1}{2}\nabla^\mu T_{\mu\nu}^{(m)}=0$. That is to say, under minimal geometry-matter coupling with an isolated $\mathscr{L}_m$ in the total Lagrangian density, the matter tensor $T^{\text{(m)}}_{\mu\nu}$ in Eq.(\ref{continuity conservation I}) has been defined in a \textit{practical} way so that $T^{\text{(m)}}_{\mu\nu}$ is automatically symmetric, Noether compatible, and covariant invariant, which naturally guarantees the local conservation $\nabla^\mu T_{\mu\nu}^{(m)}=0$. In this sense, one can regard the vanishing divergence $\nabla^\mu \left(G_{\mu\nu}+H_{\mu\nu}^{\text{(G)}}+H_{\mu\nu}^{\text{(NC)}}+H_{\mu\nu}^{(\phi)}\right)=0$ for Eq.(\ref{Field eqn}) to either \textit{imply} or \textit{confirm} the conservation $\nabla^\mu T_{\mu\nu}^{(m)}=0$.

One should be aware that in the presence of nonminimal gravity-matter couplings, like $\mathcal{R}\cdot\mathscr{L}_m$ or more generally $F\left(R,\cdots, \mathcal{R}\right) \cdot \mathscr{L}_m$ in the total Lagrangian density, the divergence $\nabla^\mu T_{\mu\nu}^{(m)}$ becomes nonzero as well and obeys the relation $\nabla^\mu T_{\mu\nu}^{(m)}= F\left(R,\cdots, \mathcal{R}\right)^{-1}
\cdot \left(\mathscr{L}_m\, g_{\mu\nu} - T_{\mu\nu}^{(m)}\right)\cdot \nabla^\mu F\left(R,\cdots, \mathcal{R}\right)$ instead \cite{Nonminimal Coupling I, Nonminimal Coupling II, AA Tian Paper}, which recovers the local conservation $\nabla^\mu T^{(m)}_{\mu\nu}=0$ for $F\left(R,\cdots, \mathcal{R}\right)=\text{constant}$.

Also, at a more fundamental level, the $T^{\text{(m)}}_{\mu\nu}$ in Eq.(\ref{continuity conservation I}) for GR and modified gravities, though practical with all desired properties, is not defined from the first-principle approach, i.e. directly from symmetry and Noether's theorem in the classical field theory. In this larger framework, the $T^{\text{(m)}}_{\mu\nu}$ in Eq.(\ref{continuity conservation I}) is often referred to as the Hilbert energy-momentum tensor: it symmetrizes the canonical energy-momentum tensor of translational invariance by adding a superpotential term, and it is a special case of the Belinfante energy-momentum tensor that minimally couples to gravity \cite{Ortin Gravity and Strings}.

\section{Nondynamical massless scalar field}\label{Sec Nondynamical scalar field}

\subsection{Nondynamical massive scalar field}\label{Sec Nondynamical massive scalar field}
\vspace{2mm}

Due to the $\lambda(\phi)$-dependence in $\mathcal{S}_\phi$, its Lagrangian density becomes $\mathscr{L}_{\phi} = -V(\phi)$  when $\lambda(\phi)\equiv 0$; considering that $V(\phi)$ is usually related to the mass of the scalar field in cosmology and high energy physics,  we will call $\phi(x^\alpha)$ nondynamical and massive for the situation $ \lambda(\phi)\equiv 0$ and $V(\phi)\neq 0 $. As such, instead of producing a prorogation equation $\Box\phi$, the extremization $\delta S/\delta \phi=0$ leads to the following constraint for  the potential $V(\phi)$:
\begin{equation}\label{nondynamical massive phi constraint}
\begin{split}
V_\phi=\widehat{f}\left(R,\cdots, \mathcal{R}  \right) \cdot h_\phi\,.
\end{split}
\end{equation}
In the meantime, Eqs.(\ref{Hmunu phi}), (\ref{Divergence phi 1}), and (\ref{phi wave 1}) reduce to become
\begin{equation}\label{nondynamical massive phi divergence}
\begin{split}
H_{\mu\nu}^{(\phi)}= \frac{1}{2}V(\phi)\,g_{\mu\nu} \qquad\text{and}\qquad
\nabla^\mu H_{\mu\nu}^{(\phi)}=\frac{1}{2}V_\phi \nabla_\nu \phi= \frac{1}{2}\widehat{f}\left(R,\cdots, \mathcal{R}  \right) \cdot h_\phi\nabla_\nu\phi \,.
\end{split}
\end{equation}
Thus, for a nondynamical yet massive scalar field, $\nabla^\mu H_{\mu\nu}^{(\phi)}$ can still balance the nontrivial divergence $\nabla^\mu H_{\mu\nu}^{(\text{NC})}$ of the nonminimal $\phi(x^\alpha)$-curvature coupling effect, while the potential or the mass of the scalar field is  restricted by the condition Eq.(\ref{nondynamical massive phi constraint}).

\subsection{Nondynamical massless scalar field}\label{Sec Nondynamical massless scalar field}
\vspace{3mm}

Within the situation $\lambda(\phi)\equiv0$, it becomes even more interesting when the potential vanishes as well in Eqs.(\ref{Lagrangian phi}), (\ref{Hmunu phi}), (\ref{Divergence phi 1}), and (\ref{phi wave 1}); we will call the scalar field \textit{nondynamical and massless}\footnote{We simply use ``massive'' and ``massless'' to distinguish the situation $V(\phi)\neq 0$ from $V(\phi)=0$ when the scalar field is nondynamical. However, we do not follow this usage to call $\phi(x^\alpha)$ ``dynamical and massless'' when $\{\lambda(\phi)\neq0, V(\phi)=0\}$, as it sounds inappropriate to from the spirit of relativity.}  for $\lambda(\phi)=0=V(\phi)$. With $\mathscr{L}_{\phi} =0$,  the total action simplifies into
\begin{eqnarray}\label{Action nondynamical massless phi}
\mathcal{S}=\int d^4x \sqrt{-g}\,\Big(R+\mathscr{L}_{\text G}+\mathscr{L}_{\text{NC}}+ 16\pi G\mathscr{L}_m  \Big)\,.
\end{eqnarray}
Since $H_{\mu\nu}^{(\phi)}=0$ and $\nabla^\mu H_{\mu\nu}^{(\phi)}=0$, the divergence $\nabla^\mu  H_{\mu\nu}^{\text{(NC)}}$ for the nonminimal coupling part as  in Eq.(\ref{Divergence NC}) can no longer be neutralized. Instead,  with $\nabla^\mu G_{\mu\nu}=0$, the generalized contracted Bianchi identities Eq.(\ref{Generalized contracted Bianchi identities}), and the covariant conservation $\nabla^\mu T_{\mu\nu}^{(m)}=0$ under minimal geometry-mater coupling, the contravariant derivative of the field equation
$G_{\mu\nu}+H_{\mu\nu}^{\text{(G)}}+H_{\mu\nu}^{\text{(NC)}}= 8\pi GT_{\mu\nu}^{(m)} $
forces $\nabla^\mu  H_{\mu\nu}^{\text{(NC)}}$  to vanish.   Together with  Eq.(\ref{Divergence NC}), this implies that to be a solution to the gravity of Eq.(\ref{Action nondynamical massless phi}), the metric tensor $g_{\mu\nu}$ must satisfy the constraint
\begin{eqnarray}\label{constraint}
\widehat{f}\left(R,\cdots, \mathcal{R}  \right) \equiv 0 \quad\text{for}\quad\phi(x^\alpha)\neq\text{constant}\,.
\end{eqnarray}
Since the nonzero divergence $\nabla^\mu H_{\mu\nu}^{\text{(NC)}}=-\frac{1}{2}\widehat{f}\left(R,\cdots, \mathcal{R}  \right) \cdot h_\phi\partial_\nu \phi$ measures the failure of diffeomorphism invariance in the componential action $\mathcal{S}_{\text{NC}}$, the consistency condition Eq.(\ref{constraint}) indicates that the symmetry breaking of diffeomorphism invariance is suppressed in gravitational dynamics of Eq.(\ref{Action nondynamical massless phi}).

Here one should note that the variation $\delta S/\delta \phi=0$ yields the condition $\widehat{f}\left(R,\cdots, \mathcal{R}  \right) \cdot  h_\phi= 0$, which also leads to  $\widehat{f}\left(R,\cdots, \mathcal{R}  \right) \equiv 0$ if the scalar field is nonconstant. In addition, the constraint $\widehat{f}\left(R,\cdots, \mathcal{R}  \right)\equiv 0$ does not mean $H_{\mu\nu}^{\text{(NC)}}=0$ or  the removal of  $\mathscr{L}_{\text{NC}}$ from  the action Eq.(\ref{Action nondynamical massless phi}). This can be seen by an analogous situation in GR: all vacuum solutions of Einstein's equation have to satisfy the condition $R\equiv 0$, but the GR action $\mathcal{S}=\int d^4x\sqrt{-g}\,\left(R+16\pi G \mathscr{L}_m  \right)$ still holds in its standard form.

After $\mathscr{L}_{\text G}$ and $\mathscr{L}_{\text{NC}} $ get specified in Eq.(\ref{Action nondynamical massless phi}), how can we know whether it yields a viable theory or not? In accordance with Eq.(\ref{constraint}), we adopt the following basic assessment.\\

\noindent \textbf{Primary test}: For the action Eq.(\ref{Action nondynamical massless phi}) to be a viable modified gravity or effective dark energy carrying a nondynamical and massless scalar field, an elementary requirement is that the function $\widehat{f}\left(R,\cdots, \mathcal{R}  \right)$ in $\mathscr{L}_{\text{NC}}$ vanishes identically for the flat and accelerating Friedmann-Robertson-Walker (FRW) Universe with the metric
\begin{equation} \label{Flat FRW}
ds^2=-dt^2+a(t)^2\sum_{i=1}^3 \left(dx^i\right)^2\quad\text{and}\quad  \ddot{a}(t)>0\,,
\end{equation}
where $a(t)$ is the cosmic scale factor and the overdot means derivative with respect to the comoving time.\\

This primary test is inspired by the fact that the observable Universe is homogeneous and isotropic at the largest cosmological scale, and the  discovery that the Universe is nearly perfectly flat and currently undergoing accelerated spatial expansion. These features have been extensively examined and received strong support from the surveys on the large scale structures, the expansion history, and the structure-growth rate of the Universe, such as the measurements of the distance modulus of Type Ia supernovae, peaks of the baryon acoustic oscillation, and temperature polarizations of the cosmic microwave background. Clearly, the primary test is updatable and subject to the progress in observational cosmology.

\subsection{Weyl dark energy}
\vspace{3mm}

Following the primary test above, one can start to explore possible modifications of GR into the total Lagrangian density $\mathscr{L}=R+f(R,\cdots,\mathcal{R})+h(\phi)\widehat{f}(R,\cdots,\mathcal{R})+16\pi G\mathscr{L}_m$ and then check the consistency condition $\widehat{f}(R,\cdots,\mathcal{R})\equiv0$ under the flat FRW metric Eq.(\ref{Flat FRW}). In the integrand of the Hilbert-Einstein action for GR, the Ricci scalar $R$ is the simplest curvature invariant formed by second-order derivatives of the metric; similarly, we can start with the simplest situation that $\widehat{f}\left(R,\cdots, \mathcal{R}  \right)$ is some quadratic Riemannian scalar. One possible example is the square
of the conformal Weyl tensor $C_{\alpha\beta\gamma\delta}=R_{\alpha\beta\gamma\delta}+\frac{1}{2}\Big( g_{\alpha\delta}R_{\beta\gamma}-g_{\alpha\gamma}R_{\beta\delta}+
g_{\beta\gamma}R_{\alpha\delta}- g_{\beta\delta}R_{\alpha\gamma} \Big)
+\frac{1}{6}\,\Big( g_{\alpha\gamma} g_{\beta\delta}
- g_{\alpha\delta} g_{\beta\gamma}  \Big)\,R$, which is the totally traceless part in the Ricci decomposition of the Riemann tensor. In this case, we consider the action
\begin{equation}\label{WEyl dark energy I}
\begin{split}
\mathcal{S}_{\mathcal{C}^2}=\int d^4x\sqrt{-g}\,\left(R+\gammaup \phi\, \mathcal{C}^2 + 16\pi G\mathscr{L}_m \right)\,,
\end{split}
\end{equation}
where $\gammaup\neq0$ is a coupling constant, and
\begin{equation}
\begin{split}
\mathcal{C}^2\coloneqq C_{\alpha\mu\beta\nu}C^{\alpha\mu\beta\nu}\equiv
\frac{1}{3}R^2-2R_{\mu\nu}R^{\mu\nu}+R_{\mu\alpha\nu\beta}R^{\mu\alpha\nu\beta}\,.
\end{split}
\end{equation}
It is straightforward to verify that $\mathcal{C}^2\equiv 0$ for arbitrary forms of the scale factor $a(t)$ in the flat FRW metric.
We would like to dub Eq.(\ref{WEyl dark energy I}) as the ``Weyl dark energy'' or ``conformal dark energy''. The field equation is $R_{\mu\nu} -\frac{1}{2}  R  g_{\mu\nu}+\gammaup H_{\mu\nu}^{(\mathcal{C}^2)}=8\pi GT_{\mu\nu}^{(m)}$, where
\begin{align}
H_{\mu\nu}^{(\mathcal{C}^2)}= &-\frac{1}{2}\phi\mathcal{C}^2 g_{\mu\nu}+
2\phi\left(\frac{1}{3} RR_{\mu\nu}
-2 R_\mu^{\;\;\,\alpha}R_{\alpha\nu}+ 
R_{\mu\alpha\beta\gamma}R_{\nu}^{\;\;\,\alpha\beta\gamma}\right)
+\frac{2}{3} \left(g_{\mu\nu}\Box-\nabla_\mu\!\nabla_\nu\right)\,\left(\phi R\right)
-2\Box \left(\phi R_{\mu\nu}\right)\nonumber\\
&+2 \nabla_\alpha\!\nabla_{\nu} \left(\phi R_{\mu}^{\;\;\alpha}\right)
+2\nabla_\alpha\!\nabla_{\mu} \left(\phi R_{\nu}^{\;\;\alpha}\right) -2 g_{\mu\nu} \nabla_\alpha\!\nabla_\beta \left(\phi  R^{\alpha\beta}  \right)
+4 \nabla^\beta \nabla^\alpha \left(\phi  R_{\alpha\mu \beta\nu} \right)\,,
\end{align}
and according to Eq.(\ref{Divergence NC}), its covariant divergence is
\begin{equation}
\begin{split}
\nabla^\mu  H_{\mu\nu}^{(\mathcal{C}^2)}= -\frac{\gammaup}{2} \mathcal{C}^2\, \nabla_\nu \phi\,.
\end{split}
\end{equation}
The Weyl dark energy $\mathcal{S}_{\mathcal{C}^2}$, where the scalar field is nondynamical and massless, can be generalized into the dynamical case
\begin{equation}\label{WEyl dark energy II}
\begin{split}
\mathcal S=\int d^4x \sqrt{-g}\,\bigg(R+h(\phi) \mathcal{C}^2 -\lambda(\phi)\cdot\nabla_\alpha\phi\nabla^\alpha\phi -V(\phi)+ 16\pi G\mathscr{L}_m \bigg)\,,
\end{split}
\end{equation}
for which the constraint $\mathcal{C}^2\equiv 0$ is no longer necessary and should be removed.

The complete validity of the Weyl dark energy $\mathcal{S}_{\mathcal{C}^2}$ or its extension  Eq.(\ref{WEyl dark energy II}), including the value of the coupling strength $\gammaup $ in  $\mathcal{S}_{\mathcal{C}^2}$, should be carefully constrained by the observational data from astronomical surveys. Following the field equation of  $\mathcal{S}_{\mathcal{C}^2}$, consider a $\mathcal{C}^2$CDM model (i.e. $\mathcal{C}^2$ cold dark matter) for the Universe instead of $\Lambda$CDM. Then the first Friedmann equation under the flat FRW metric reads
\begin{equation}\label{Friedmann equation Gravity form}
\begin{split}
H^2\,=\,\frac{8}{3}\pi G\left[\rho_{M0}\left(\frac{a_0}{a}\right)^3+\rho_{r0}\left(\frac{a_0}{a}\right)^4
+\rho_{\mathcal{C}^2}  \right]\,,
\end{split}
\end{equation}
where the densities of nonrelativistic matter $\rho_M (t)$ and relativistic matter $\rho_r (t)$ have been related to their present-day values $\rho_{M0}$ and $\rho_{r0}$ via by the continuity equation $\dot\rho+3H\rho(1+w)=0$, with the equation of state parameters being $w_M=0$ and $w_r=1/3$, respectively. Also, $H\coloneqq \dot a/a$ is the evolutionary Hubble parameter, and $\rho_{\mathcal{C}^2}$ denotes the effective energy density of the Weyl dark energy,
\begin{equation}
\begin{split}
\rho_{\mathcal{C}^2}=\gammaup\,[ & 5\dot{\phi}\frac{\ddot{a}}{a}\frac{\dot{a}}{a}
-2\dot{\phi}\frac{\ddot{a}^2}{a^2}\frac{\dot{a}}{a}
-\dot{\phi}\frac{\dot{a}^3}{a^3}-2\dot{\phi}\frac{\dot{a}^5}{a^5}
+5\phi \frac{\dddot{a}}{a}\frac{\dot{a}}{a} +2\phi\frac{\ddot{a}^3}{a^3}\\
-  &4\phi \frac{\ddot{a}}{a}\frac{\dot{a}^2}{a^2}
 -4\phi \frac{\dddot{a}}{a}\frac{\ddot{a}}{a}\frac{\dot{a}}{a}
+4\phi \frac{\ddot{a}^2}{a^2}\frac{\dot{a}^2}{a^2}
-6\phi \frac{\ddot{a}}{a}\frac{\dot{a}^4}{a^4} +8\phi \frac{\dot{a}^6}{a^6}
\Big]\,.
\end{split}
\end{equation}
Employing the cosmological redshift $z\coloneqq a_0/a-1$ as well as the replacements $\ddot a/a=\dot H +H^2$ and $\dddot a/a=\ddot H+3 \dot H H+H^3$, Eq.(\ref{Friedmann equation Gravity form}) can be parameterized into
\begin{equation}\label{Friedmann equation Astrophysics form}
\begin{split}
H(z; H_0, \bm{p}) = H_0\sqrt{\Omega_{M0} (1+z)^3 + \Omega_{r0} (1+z)^4  + \Omega_{\mathcal{C}^2}}\,,
\end{split}
\end{equation}
where $H_0$ represents the Hubble constant $H(z=0)$,  $\Omega_{M0}=8\pi G\rho_{M0}/(3H_0^2)$, $\Omega_{r0}=8\pi G\rho_{r0}/(3H_0^2)$, and
\begin{align}
\Omega_{\mathcal{C}^2}=\frac{32\pi G}{H_0^2}\gammaup\Bigg\{&\dot{\phi}H\Big[5\left(\dot H +H^2\right)
-2\left(\dot H +H^2\right)^2 - H^2-2H^4\Big]+
\phi H\left(5-4\dot H -4H^2\right)\left(\ddot H+3\dot H H+H^3\right)\nonumber\\
&+ 8\phi H^6+\phi\left(\dot H +H^2\right)  \Big[\left(2\dot H +2H^2 +4H^2\right)\left(\dot H +H^2\right)-4H^2-6H^4 \Big]\Bigg\}\,.
\end{align}
Typically, we can use the Markov-Chain Monte-Carlo engine CosmoMC \cite{CosmoMC}  to explore the parameter space $\bm{p}=(\Omega_{M0}, \Omega_{r0},\gammaup)$ for the Weyl dark energy $\mathcal{S}_{\mathcal{C}^2}$,  and find out how well  it matches the various sets of observational data. This goes beyond the scope of this paper and will be analyzed separately.



\section{Applications}\label{Sec Applications}

\subsection{Chern-Simons gravity}\label{Sec Chern-Simons gravity}

The four-dimensional Chern-Simons modification of GR was proposed by the action \cite{Chern-Simons 1} (note that not to confuse with the traditional gauge gravity carrying a three-dimensional Chern-Simons term \cite{3D Chern-Simons})
\begin{equation}\label{Chern-Smons nondynamical}
\begin{split}
\mathcal{S}_{\text{CS}}=\int d^4x\sqrt{-g}\,\left(R+\gammaup \phi \frac{{}^*RR}{\sqrt{-g}} + 16\pi G\mathscr{L}_m \right)\,.
\end{split}
\end{equation}
The scalar field $\phi=\phi(x^\alpha)$ is nonminimally coupled to the Chern-Pontryagin density
${}^*RR  \coloneqq
{}^{*}R_{\alpha\beta\gamma\delta} R^{\alpha\beta\gamma\delta}= \frac{1}{2} \epsilon_{\alpha\beta\mu\nu} R^{\mu\nu}_{\;\;\;\;\,\gamma\delta}R^{\alpha\beta\gamma\delta}$,
where
${}^{*}R_{\alpha\beta\gamma\delta}\coloneqq\frac{1}{2}\epsilon_{\alpha\beta\mu\nu} R^{\mu\nu}_{\;\;\;\;\gamma\delta}$ is the left dual of the Riemann tensor, and $\epsilon_{\alpha\beta\mu\nu}$
represents the totally antisymmetric Levi-Civita pseudotensor with $\epsilon_{0123}=\sqrt{-g}$ and $\epsilon^{0123}=1/\sqrt{-g}$. The field equation reads $R_{\mu\nu} -\frac{1}{2}  R  g_{\mu\nu}
 + \gammaup H_{\mu\nu}^{\text{(CP)}}=8\pi G T_{\mu\nu}^{\text{(m)}}$,
where $H_{\mu\nu}^{\text{(CP)}}\cong\frac{1}{\sqrt{-g}} \frac{\delta \left(\phi {}^*RR\right)}{\delta g^{\mu\nu}}$ collects the contributions from the $\phi(x^\alpha)$-coupled Chern-Pontryagin density,
\begin{equation}\label{LBD field CP}
\begin{split}
\sqrt{-g}\,H_{\mu\nu}^{\text{(CP)}}=\, 2\partial^\xi\phi\cdot
\left( \epsilon_{\xi\mu\alpha\beta}\nabla^\alpha R^\beta_{\;\;\nu} +
\epsilon_{\xi\nu\alpha\beta}\nabla^\alpha R^\beta_{\;\;\mu} \right)+
2\partial_\alpha \partial_\beta  \phi \cdot \left({}^*R^{\alpha\;\;\,\beta}_{\;\;\,\mu\;\;\nu} +{}^*R^{\alpha\;\;\,\beta}_{\;\;\,\nu\;\;\mu}\right)\,.
\end{split}
\end{equation}

According to the general theory in Secs.~\ref{Sec Divergence of field equation} and ~\ref{Sec Nondynamical massless scalar field}, the Chern-Simons gravity
Eq.(\ref{Chern-Smons nondynamical}) involves a nondynamical and massless scalar field. Identifying $\widehat{f}\left(R,\cdots, \mathcal{R}  \right)$ as ${}^*RR/\sqrt{-g}$ and with $h(\phi)=\gammaup\phi$ in  Eqs.(\ref{Divergence NC}) and (\ref{constraint}), we obtain the divergence
\begin{equation}
\nabla^\mu  H_{\mu\nu}^{\text{(CP)}}
=-\frac{\gammaup {}^*RR}{2\sqrt{-g}}\cdot \partial_\nu \phi\,,
\end{equation}
as well as the constraint ${}^*RR\equiv 0$
for nontrivial $\phi(x^\alpha)$. It can be easily verified that ${}^*RR$ vanishes for the flat and accelerating FRW Universe, and thus passes the primary test in Sec.~\ref{Sec Nondynamical massless scalar field}. Also the condition ${}^*RR\equiv 0$ only applies to the action Eq.(\ref{Chern-Smons nondynamical}), and is invalid for the massive Chern-Simons gravity $\mathscr{L}= R+\gammaup \phi \frac{{}^*RR}{\sqrt{-g}} -V(\phi)+ 16\pi G\mathscr{L}_m $ or the dynamical case
$\mathscr{L}= R+\gammaup \phi  \frac{{}^*RR}{\sqrt{-g}} -\lambda(\phi)\cdot\nabla_\alpha\phi\nabla^\alpha\phi+ 16\pi G\mathscr{L}_m$.


\subsection{Reduced Gauss-Bonnet dark energy}\label{Sec Gauss-Bonnet dark energy}
\vspace{3mm}

The Gauss-Bonnet dark energy was introduced by the action
$\mathcal S_{\text{GB}}^{(1)}=\int d^4x \sqrt{-g}\,
\bigg(R+h(\phi) \mathcal{G}-\bar\lambdaup\nabla_\alpha\phi\nabla^\alpha\phi-V(\phi)+ 16\pi G\mathscr{L}_m\bigg)$ \cite{GaussBonnet dark energy},
where $\bar\lambda\in\left\{\pm1, 0\right\}$, and the scalar field is nonminimally coupled to the Gauss-Bonnet invariant
$\mathcal{G} \coloneqq \left(\frac{1}{2}\epsilon_{\alpha\beta\gamma\zeta}R^{\gamma\zeta\eta\xi}\right)
\cdot \left(\frac{1}{2}\epsilon_{\eta\xi\rho\sigma} R^{\rho\sigma\alpha\beta}\right)
\equiv R^2-4R_{\alpha\beta}R^{\alpha\beta}+R_{\alpha\mu\beta\nu}R^{\alpha\mu\beta\nu}$.
If $\phi(x^\alpha)$ is nondynamical with $\bar\lambdaup=0$, the action $\mathcal S_{\text{GB}}^{(1)}$ reduces to become $\mathcal S_{\text{GB}}^{(2)}=\int d^4x \sqrt{-g}\,\bigg(R+h(\phi) \mathcal{G} -V(\phi)+ 16\pi G\mathscr{L}_m\bigg)$, and according to Eq.(\ref{nondynamical massive phi constraint}) with $\widehat{f}\left(R,\cdots, \mathcal{R}  \right)$ identified as the Gauss-Bonnet invariant, $V(\phi)$ has to satisfy the constraint $V_\phi=\mathcal{G}\,h_\phi$.
Moreover, the nonminimally coupled $h(\phi) \mathcal{G}$ part in $\mathcal S_{\text{GB}}^{(1)}$ and $\mathcal S_{\text{GB}}^{(2)}$ contributes to the field equation by
\begin{align}
H_{\mu\nu}^{\text{(GB)}}=\,
&2R\left(g_{\mu\nu}\Box
-\nabla_\mu\!\nabla_\nu\right)h
+4R_{\mu}^{\;\;\,\alpha}\nabla_\alpha\!\nabla_{\nu}h +
4R_{\nu}^{\;\;\,\alpha}\nabla_\alpha\!\nabla_{\mu} h \nonumber\\
&-4R_{\mu\nu}\Box h -4g_{\mu\nu} \cdot R^{\alpha\beta}\nabla_\alpha\!\nabla_\beta h+4R_{\alpha\mu \beta\nu}
\nabla^\beta \nabla^\alpha  h\,, \label{GB variation}
\end{align}
where, compared with the original field equation in Ref.\cite{GaussBonnet dark energy}, we have removed the algebraic terms in $H_{\mu\nu}^{\text{(GB)}}$ by the Bach-Lanczos identity $\frac{1}{2}\mathcal{G}g_{\mu\nu}\equiv
2 RR_{\mu\nu}-4 R_\mu^{\;\;\,\alpha}R_{\alpha\nu}-4 R_{\alpha\mu\beta\nu}R^{\alpha\beta}
+2R_{\mu\alpha\beta\gamma}R_{\nu}^{\;\;\,\alpha\beta\gamma}$ \cite{EMC LBD gravity}.
The divergence of $H_{\mu\nu}^{\text{(GB)}}$, in accordance with Eq.(\ref{Divergence NC}), reads
\begin{equation}
\nabla^\mu  H_{\mu\nu}^{\text{(GB)}}
=-\frac{1}{2}\,\mathcal{G}\cdot h_\phi\partial_\nu \phi \,.
\end{equation}
However, it would be problematic if one further reduces  $\mathcal S_{\text{GB}}^{(2)}$ into
\begin{equation}
\mathcal S_{\text{GB}}^{(3)}=\int d^4x \sqrt{-g}\,\bigg(R+h(\phi) \mathcal{G}+ 16\pi G\mathscr{L}_m\bigg)\,,
\end{equation}
where $\phi(x^\alpha)$ is both nondynamical and massless. The metric tensor has to satisfy $\mathcal{G}\equiv 0$
to be a solution to the field equation $R_{\mu\nu} -\frac{1}{2}  R  g_{\mu\nu}
+ H_{\mu\nu}^{\text{(GB)}} =8\pi G T_{\mu\nu}^{\text{(m)}}$ for the reduced Gauss-Bonnet dark energy $S_{\text{GB}}^{(3)}$. For the flat FRW Universe with the metric Eq.(\ref{Flat FRW}), the Gauss-Bonnet invariant is
\begin{equation}
\begin{split}
\mathcal{G}=24\frac{\dot{a}^2\ddot{a}}{a^3}\,,
\end{split}
\end{equation}
and thus $\mathcal{G}$ vanishes only if the Universe were of static state ($\dot a=0$) or constant acceleration ($\ddot a=0$). Hence, the constraint $\mathcal G\equiv 0$ for $S_{\text{GB}}^{(3)}$  is inconsistent with the cosmic acceleration, which indicates that unlike $S_{\text{GB}}^{(1)}$ and $S_{\text{GB}}^{(2)}$,  $S_{\text{GB}}^{(3)}$ is oversimplified and can not be a viable candidate of effective dark energy.

\subsection{Generalized scalar-tensor theory}\label{Sec Generalized scalar-tensor gravity}
\vspace{3mm}

Since $\mathcal{S}_{\text{HE}}$ and $\mathcal{S}_{\text{G}}$ in Eq.(\ref{Generic action}) respect the diffeomorphism invariance and the (generalized) contracted Bianchi identities, in this subsection we will ignore them and focus on the following scalar-tensor-type gravity in the Jordan frame:
\begin{equation}\label{Generalized ST}
\begin{split}
\mathcal{S}_{\text{ST}}\,
&= \int d^4x \sqrt{-g}\,\bigg(f(R, \phi)+ \mathscr{L}_{\text{NC}} + \mathscr{L}_\phi  +16\pi G\mathscr{L}_m \bigg)\\
&=\int d^4x \sqrt{-g}\,\bigg(f(R, \phi)+ h(\phi)\cdot \widehat{f}\left(R,\cdots, \mathcal{R}  \right)- \lambda(\phi)\cdot\nabla_\alpha\phi\nabla^\alpha\phi  -V(\phi)  +16\pi G\mathscr{L}_m\bigg)\,,
\end{split}
\end{equation}
where $f(R,\phi)$ is a hybrid function of the Ricci scalar and the scalar field. $f(R,\phi)$ contributes to the field equation by
\begin{equation}
\begin{split}
H_{\mu\nu}^{f(R,\phi)}\,\cong\frac{1}{\sqrt{-g}}\frac{\delta\left(\!\!\sqrt{-g}\,f(R,\phi)\right)}{\delta g^{\mu\nu}}=-\frac{1}{2}f(R,\phi)\cdot g_{\mu\nu}+f_R  R_{\mu\nu}
+\left(g_{\mu\nu}\Box-\nabla_\mu\nabla_\nu  \right)f_R \,,
\end{split}
\end{equation}
where  $f_R=f_R(R,\phi)=\partial f(R,\phi)/\partial R$.
With the Bianchi identity $\nabla^\mu  \left(R_{\mu\nu} -\frac{1}{2}  R
g_{\mu\nu}\right)=0$ and the third-order-derivative commutator
$ \left(\nabla_\nu\Box -\Box\nabla_\nu \right) f_R =
-R_{\mu\nu}\nabla^\mu
f_R$, explicit calculations yield
\begin{equation}\label{Divergence fRphi}
\begin{split}
\nabla^\mu H_{\mu\nu}^{f(R,\phi)}\,= -\frac{1}{2}f_\phi\cdot\nabla_\nu \phi  \,,
\end{split}
\end{equation}
where  $f_\phi=f_\phi(R,\phi)=\partial f(R,\phi)/\partial \phi$.
On the other hand, the kinematical wave equation $\delta\mathcal{S}_{\text{ST}}/\delta\phi=0$ reads
$
2\lambda(\phi)\cdot\Box\phi= -f_\phi- \widehat{f}\cdot h_\phi- \lambda_\phi\cdot\nabla_\alpha\phi\nabla^\alpha\phi + V_\phi
$,
which recasts the divergence $\nabla^\mu H_{\mu\nu}^{(\phi)}=-
\frac{1}{2}\left(\lambda_\phi\cdot\nabla_\alpha\phi\nabla^\alpha\phi+2\lambda(\phi)\cdot\Box\phi -V_\phi\right)\cdot\nabla_\nu \phi$ as in Eq.(\ref{Divergence phi 1}) into
\begin{equation}\label{Divergence phi 3}
\begin{split}
\nabla^\mu H_{\mu\nu}^{(\phi)}= \frac{1}{2}\bigg(f_\phi+\widehat{f}\left(R,\cdots, \mathcal{R} \right)\cdot h_\phi\bigg) \cdot \nabla_\nu\phi \,.
\end{split}
\end{equation}
Hence, with Eqs.(\ref{Divergence NC}), (\ref{Divergence fRphi}) and(\ref{Divergence phi 3}),
we immediately learn that the field equation $H_{\mu\nu}^{f(R,\phi)}+H_{\mu\nu}^{\text{(NC)}}+H_{\mu\nu}^{(\phi)}= 8\pi GT_{\mu\nu}^{(m)}$ for the scalar-tensor-type gravity $\mathcal{S}_{\text{ST}}$ is divergence free. By the total Lagrangian density for the sake of simplicity, the concretization of Eq.(\ref{Generalized ST}) includes, for example, standard Brans-Dicke gravity
$\mathscr{L}=\phi R-\frac{\omega_{\text{BD}}}{\phi}\,
\nabla_{\alpha}\phi \nabla^{\alpha} \phi
+16\pi \mathscr{L}_m$ (where Newton's constant $G$ is encoded into $\phi^{-1}$ in the spirit of Mach's principle) \cite{Brans Dicke},
generalized Brans-Dicke gravity $\mathscr{L}=\phi R-\frac{\omega(\phi)}{\phi} \,\nabla_{\alpha}\phi \nabla^{\alpha}\phi-V(\phi)+16\pi G \mathscr{L}
_m $  with a self-interaction potential, Lovelock-Brans-Dicke gravity $\mathscr{L}=\phi\left(R
+\frac{a}{\sqrt{-g}} {}^*RR + b\mathcal{G}\right)
 -\frac{\omega_{\text{L}}}{\phi}\nabla_\alpha \phi \nabla^\alpha\phi-V(\phi) +16\pi G \mathscr{L}_m$ \cite{EMC LBD gravity},  Lovelock-scalar-tensor gravity $\mathscr{L}=f_1(\phi) R
+\frac{f_2(\phi)}{\sqrt{-g}} {}^*RR
+ f_3(\phi)\mathcal{G}
-\omega(\phi)\cdot\nabla_\alpha \phi \nabla^\alpha\phi
-V(\phi) +16\pi G \mathscr{L}_m$ \cite{EMC LBD gravity}, minimal dilatonic gravity $\mathscr{L}=\phi R-2\Lambda U(\phi)$ \cite{Fiziev minimal dilatonic gravity}, Gauss-Bonnet dilatonic gravity $\mathscr{L}= R-\nabla_\alpha\phi\nabla^\alpha\phi
+e^{-\gamma\phi}\mathcal{G}$ or $\mathscr{L}= e^{-\gamma\phi}(R-\nabla_\alpha\phi\nabla^\alpha\phi
+\mathcal{G})$ motivated by the low-energy heterotic string theory \cite{Gauss-Bonnet dilaton}, and the standard scalar-tensor gravity $\mathscr{L}=F(\phi)R-Z(\phi) \cdot\nabla_{\alpha}\phi \nabla^{\alpha} -V(\phi) +16\pi G \mathscr{L}_m $ \cite{Scalar Tensor Theory};  all these examples satisfy the local energy-momentum conservation $\nabla^\mu T_{\mu\nu}^{(m)}=0$ and have divergence-free field equations.

\subsection{Hybrid metric-Palatini $f(R)$ gravity}
\vspace{3mm}

So far we have been using the metric formulation for the curvature invariants; however, the local conservation can be proved for the Palatini or hybrid metric-Palatini $f(R)$ gravity without referring to the Palatini formulation of the (generalized) Bianchi identities.
Consider the following hybrid metric-Palatini $f(R)$ action
\begin{equation}
\begin{split}
\mathcal{S}_{\text Hf}^{(1)}\,=\int d^4x \sqrt{-g}\,\bigg(R+ f(\hat{R}) +16\pi G\mathscr{L}_m\bigg)\,,
\end{split}
\end{equation}
where $R$ is the usual Ricci scalar for the metric $g_{\mu\nu}$, while $\hat R=\hat R(\bm g, \hat{\bm\Gamma})=g^{\mu\nu}\hat{R}_{\mu\nu}(\hat{\bm\Gamma})$ denotes the Palatini Ricci scalar, with the Palatini Ricci tensor given by $\hat{R}_{\mu\nu}(\hat{\bm\Gamma})
=\hat{R}^\alpha_{\;\;\mu\alpha\nu}(\hat{\bm\Gamma})=\partial_\alpha \hat{\Gamma}^\alpha_{\nu\mu}-\partial_\nu \hat{\Gamma}^\alpha_{\alpha\mu}+\hat{\Gamma}^\alpha_{\alpha\zeta}
\hat{\Gamma}^\zeta_{\mu\nu}
-\hat{\Gamma}^\alpha_{\mu\zeta}\hat{\Gamma}^\zeta_{\alpha\nu}$.  Variation of $\mathcal{S}_{\text Hf}$ with respect to the independent connection $\hat\Gamma^\alpha_{\mu\nu}$ yields $\hat{\nabla}_\alpha (\sqrt{-g}\, f_{\hat R}g^{\mu\nu})=0$, where $\hat{\nabla}$ is the covariant derivative of the connection and $f_{\hat{R}} \coloneqq df(\hat R)/d\hat R$. Thus, $\hat{\nabla}$ is compatible with the auxiliary metric $f_{\hat R}g_{\mu\nu}\eqqcolon \hat g_{\mu\nu}$, as $\sqrt{-\hat{g}}\,\hat{g}^{\mu\nu}=\sqrt{-g}\, f_{\hat R}g^{\mu\nu}$. Relating $\hat g_{\mu\nu}$ to $g_{\mu\nu}$ by the conformal transformation $g_{\mu\nu}\mapsto\hat g_{\mu\nu}$, and accordingly rewriting $\hat R_{\mu\nu}$ and $\hat R$ in the metric formulation, one could find that $\mathcal{S}_{\text Hf}^{(1)}$ is equivalent to \cite{Metric-Palatini gravity}
\begin{equation}
\begin{split}
\mathcal{S}_{\text Hf}^{(2)}\,=\int d^4x \sqrt{-g}\,\bigg(R+\phi R+ \frac{3}{2\phi}\nabla_\alpha\phi\nabla^\alpha\phi 
 -V(\phi)  +16\pi G\mathscr{L}_m\bigg)\,,
\end{split}
\end{equation}
where $\phi(x^\alpha)=f_{\hat R}(\hat R)$ and $V(\phi)=f_{\hat R}\hat R-f(\hat R)$.
$\mathcal{S}_{\text Hf}^{(2)}$ is just the mixture of GR and the $\omega_{\text{BD}}=-3/2$ Brans-Dicke gravity.
Recall that Eq.(\ref{Generalized ST}) has employed the generic function $f(R,\phi)$ for $\mathcal{S}_{\text{ST}}$, which includes the hybrid situations like $f(R,\phi)=R+\phi R$. Hence, following Sec.~\ref{Sec Generalized scalar-tensor gravity}, it is clear that the hybrid scalar-tensor gravity $\mathcal{S}_{\text Hf}^{(2)}$ and thus the hybrid metric-Palatini $f(R)$ gravity $\mathcal{S}_{\text Hf}^{(1)}$ have divergence-free field equations and respect the local energy-mentum conservation.



\section{Conclusions}\label{Sec Conclusions}

In this paper, we have investigated the covariant invariance of the field equation for a large class of hybrid modified gravity $\mathscr{L} =R+f\left(R,\cdots, \mathcal{R}  \right)+h(\phi)\cdot \widehat{f}\left(R,\cdots, \mathcal{R}  \right) -\lambda(\phi)\cdot\nabla_\alpha\phi\nabla^\alpha\phi  -V(\phi)+16\pi G \mathscr{L}_m$. For the four components $\mathscr{L}_{\text{HE}}=R$, $\mathscr{L}_{\text G}=f\left(R,\cdots, \mathcal{R}  \right)$, $\mathscr{L}_{\text{NC}}=h(\phi)\cdot \widehat{f}\left(R,\cdots, \mathcal{R}  \right)$, and $\mathscr{L}_{\phi}=-\lambda(\phi)\cdot\nabla_\alpha\phi\nabla^\alpha\phi  -V(\phi)$, we have calculated their contributions $\left\{G_{\mu\nu}, H_{\mu\nu}^{\text{(G)}}, H_{\mu\nu}^{\text{(NC)}}, H_{\mu\nu}^{(\phi)}\right\}$ to the gravitational field equation along with the respective divergences, which proves the divergence-freeness of the field equation (\ref{Field eqn}) and confirms/proves the local energy-momentum conservation under minimal gravity-matter coupling.

$H_{\mu\nu}^{\text{(NC)}}$ and $H_{\mu\nu}^{(\phi)}$ fail to obey the generalized contracted Bianchi identities due to the presence of the background scalar field $\phi(x^\alpha)$, but fortunately, the two nontrivial divergences $\nabla^\mu H_{\mu\nu}^{\text{(NC)}}$ and $\nabla^\mu H_{\mu\nu}^{(\phi)}$ exactly cancel out each other. When $\phi(x^\alpha)$ is nondynamical and massless, i.e. $\lambda(\phi)=0=V(\phi)$, the divergence $\nabla^\mu  H_{\mu\nu}^{\text{(NC)}}
=-\frac{1}{2} \widehat{f}\left(R,\cdots, \mathcal{R}  \right) \cdot h_\phi\partial_\nu \phi $ is forced to vanish, which implies the constraint
$\widehat{f}\left(R,\cdots, \mathcal{R}  \right)\equiv 0$ for nonconstant $\phi(x^\alpha)$.
We have suggested a primary viability test for the gravity $\mathscr{L} =R+f\left(R,\cdots, \mathcal{R}  \right)+h(\phi)\cdot \widehat{f}\left(R,\cdots, \mathcal{R}  \right) +16\pi G \mathscr{L}_m$ by requiring that $\widehat{f}\left(R,\cdots, \mathcal{R}  \right)$ vanishes identically for the flat and accelerating FRW Universe, and a simplest example is the Weyl dark energy $\mathscr{L}=R+\gammaup \phi\, \mathcal{C}^2 + 16\pi G\mathscr{L}_m $.

With the general theory developed in Secs.~\ref{Sec Divergence of field equation} and ~\ref{Sec Nondynamical massless scalar field}, we have considered the  applications to the Chern-Simons gravity, Gauss-Bonnet dark energy, and various (generalized) scalar-tensor gravities. In fact, the theory $\mathscr{L}_{\text{ST}}=f(R, \phi)+ h(\phi)\cdot \widehat{f}\left(R,\cdots, \mathcal{R}  \right)- \lambda(\phi)\cdot\nabla_\alpha\phi\nabla^\alpha\phi  -V(\phi)  +16\pi G\mathscr{L}_m$ in Sec.~\ref{Sec Generalized scalar-tensor gravity} can be further extended into $\mathscr{L}_{\text{EST}}=f\left(R,\cdots, \mathcal{R}, \phi\right)- \lambda(\phi)\cdot\nabla_\alpha\phi\nabla^\alpha\phi  -V(\phi)  +16\pi G\mathscr{L}_m$, for which we conjecture that
the covariant conservation still holds, with
\begin{equation}\label{Divergence phi final}
\begin{split}
H_{\mu\nu}^{(f)}\,\cong\frac{1}{\sqrt{-g}}\frac{\delta\left[\!\!\sqrt{-g}\,f(R,\cdots, \mathcal{R}, \phi)\right]}{\delta g^{\mu\nu}}\quad\text{and}\quad
\nabla^\mu H_{\mu\nu}^{(f)}= -\frac{1}{2} f_\phi(R,\cdots, \mathcal{R}, \phi) \cdot \nabla_\nu\phi\,.
\end{split}
\end{equation}
However, this divergence relation has not yet been proved in this paper, and we hope it could be solved in future.

In prospective studies, we will take into account the existent candidates of the energy-momentum pseudotensor $t_{\mu\nu}$ for the gravitation field (cf. Ref.\cite{Padmanabhan gravitational pseudotensor} for a review), and discuss the global conservation $\nabla^\mu \left(T_{\mu\nu}^{(m)}+t_{\mu\nu}\right)=0$.
Also, we will make use of more fundamental definitions of the  energy-momentum tensor, and look deeper into the conservation problem in modified gravities from the perspective of Noether's theorem and the classical field theory.


\end{document}